\documentclass{amsart}
\usepackage{mathptmx}
\usepackage{color}
\usepackage{amsmath}
\usepackage{graphicx}
\usepackage{xcolor} 
 \usepackage{multirow}
\usepackage{algorithmicx}
\usepackage{algorithm}
\usepackage[noend]{algpseudocode}

\begin{document}

\title[PAMPO]{PAMPO: using pattern matching and pos-tagging for effective Named Entities recognition in Portuguese}

%
\author[Concei\c c\~ao Rocha et al.]{Concei\c{c}\~ao Rocha}%

\author[]{Al\'{\i}pio M\'ario Jorge}
\author[]{Roberta Akemi Sionara}
\author[]{Paula Brito}
\author[]{Carlos Pimenta}
\author[]{Solange Oliveira Rezende}

\address{Concei\c c\~ao Rocha}
\curraddr{LIAAD/INESC TEC, Campus da FEUP, Rua Dr Roberto Frias, 4200-465 Porto}
\email{conceicao.n.rocha@inesctec.pt}
\address{Al\'{\i}pio M\'ario Jorge}
\curraddr{INESCTEC and FCUP, Rua do Campo Alegre, s/n, 4169-007 Porto}
\address{Roberta Akemi Sionara and Solange Oliveira Rezende}
\curraddr{ICMC-USP, University of S\~ao Paulo, S\~ao Carlos, Brazil}
\address{Paula Brito}
\curraddr{INESC TEC and FEP, Rua Dr Roberto Frias, 4200-464 Porto}
\address{Carlos Pimenta}
\curraddr{FEP and OBEGEF, FEP - Gabinete 519, Rua Dr. Roberto Frias, 4200-464 Porto, Portugal}




\begin{abstract}
This paper deals with the entity extraction task (named entity recognition) of a text mining process that aims at unveiling non-trivial semantic structures, such as relationships and interaction between entities or communities. In this paper we present a simple and efficient named entity extraction algorithm. The method, named PAMPO (PAttern Matching and POs tagging based algorithm for NER), relies on flexible pattern matching, part-of-speech tagging and lexical-based rules. It was developed to process texts written in Portuguese, however it is potentially applicable to other languages as well.

We compare our approach with current alternatives  that support Named Entity Recognition (NER) for content written in Portuguese. These are \emph{Alchemy}, \emph{Zemanta} and \emph{Rembrandt}. Evaluation of the efficacy of the entity extraction method on several texts written in Portuguese indicates a considerable improvement on $recall$ and $F_1$ measures. 
\end{abstract}

\maketitle

\section{Introduction}
\label{intro}
Nowadays, a large amount of information is produced and shared in unstructured form, mostly unstructured text~\cite{Campos12,Conrado14}. This information can be exploited in decision making processes but, to be useful, it should be transformed and presented in ways that make its intrinsic knowledge more readily intelligible. For that, we need efficient methods and tools that quickly extract useful information from unstructured text collections. Such demand can be observed, for instance, in Biology, where researchers, in order to be abreast of all developments, need to analyse new biomedical literature on a daily basis~\cite{Cohen05}. Another application is on fraud and corruption studies where the network information --- the set of actors and their relationships --- is implicitly stored in unstructured natural-language documents~\cite{Mooney05}. Hence, text mining and information extraction are required to pre-process the texts in order to extract the entities and the relations between them. 

Information extraction is a challenging task mainly due to the ambiguous features of natural-language. Moreover, most tools need to be adapted to different human languages and to different domains~\cite{Sagayam12}. In fact, the language of the processed texts is still the decisive factor when choosing among existing information extraction technologies. This is also true for the task of entity extraction (Named Entity Recognition - NER). 

For several reasons, text mining tools are typically first developed for English and only afterwards extended to other languages. Thus, there are still relatively few text mining tools for Portuguese and even less that are freely accessible. In particular, for the named entities recognition task in Portuguese texts, we find three extractors available: \emph{Alchemy\footnote{http://www.alchemyapi.com}}, \emph{Zemanta\footnote{http://www.zemanta.com}} and \emph{Rembrandt}~\cite{Cardoso12}. We also find some studies where the measures ($recall$, $precision$ and $F_1$) for those extractors are computed and compared~\cite{Rizzo11}, but their comparative effectiveness remains domain and final purpose dependent. 

In this work, we present PAMPO (PAttern Matching and POs tagging based algorithm for NER), a new method to automatically extract named entities from unstructured texts, applicable to the Portuguese language but potentially adaptable to other languages as well. The method relies on flexible pattern matching, part-of-speech tagging and lexical-based rules. All steps are implemented using free software and taking advantage of various existing packages.

The process has been developed using as case-study a specific book written in Portuguese, but it has since been used in other applications and successfully tested in different text collections. In this paper, we describe the evaluation procedures on independent textual collections, and produce a comparative study of PAMPO with other existing tools for NER.

\section{Related Work}
\label{sec:1}
In 1991, Lisa F. Rau presented a paper describing an algorithm, based on heuristics and handcrafted rules, to automatically extract company names from financial news~\cite{Rau91}. This was one of the first research papers on the NER field~\cite{Nadeau07}. NER was first introduced as an information extraction task but since then its use in natural language text has spread widely through several fields, namely Information Retrieval, Question Answering, Machine Translation, Text Translation, Text Clustering and Navigation Systems~\cite{Shaalan14}. In an attempt to suit the needs of each application, nowadays, a NER extraction workflow comprises not only analysing some input content and detecting named entities, but also assigning them a type and a list of URIs for disambiguation~\cite{Rizzo12}. New approaches have been developed with the application of Supervised machine Learning (SL) techniques~\cite{Rizzo11} and NER evolved to NERC --- Named Entity Recognition and Classification. The handicap of those techniques is the requirement of a training set, {\em i.e.}, a data set manually labelled. Therefore, the NER task depends also on the data set used to train the NER extraction algorithm. 

Currently, many existing approaches for NER/NERC are implemented and available as downloadable code, APIs or web applications, {\em i.e.}, as tools or services available on the web. A thorough search produces the following list: AIDA\footnote{http://www.mpi-inf.mpg.de/tago-naga/aida/}, AlchemyAPI\footnote{http://www.alchemyapi.com/api/demo.html}, Apache Stanbol\footnote{http://dev.iks-project.eu:8081/enhancer},  CiceroLite\footnote{http://demo.languagecomputer.com/cicerolite}, DBpedia Spotlight\footnote{http://dbpedia-spotlight.github.com/demo}, Evri\footnote{http://www.evri.com/developer/index.html}, Extractiv\footnote{http://extrativ.com}, FOX\footnote{http://askw.org/Projects/FOX.html}, FRED\footnote{http://wit.istc.cnr.it/stlab-tools/fred}, Lupedia\footnote{http://lupedia.ontotext.com}, NERD\footnote{http://nerd.eurecom.fr}, Open Calais\footnote{http://viewer.opencalais.com}, PoolParty Knowledge Discoverer\footnote{http://poolparty.biz/demozone/general}, Rembrandt\footnote{http://xldb.di.fc.ul.pt/Rembrandt/}, ReVerb\footnote{http://reverb.cs.washington.edu}, Saplo\footnote{http://saplo.com}, Se\-mio\-search Wikifier\footnote{http://wit.istc.cnr.it/stlab-tools/wikifier}, Wikimeta\footnote{http://www.wikimeta.com/wapi/semtag.pl}, Yahohh! Content Analysis (YCA)\footnote{http://developer.yahoo.com/search/content/V2/contentAnalysis}, Zemanta\footnote{http://www.zemanta.com/demo/}. More detailed information may be found in ~\cite{Rizzo12,Gangemi13,Mendes11,Cardoso12}, where the authors compare the services' strengths and weaknesses and compute some measures for their performance.

Nadeau {\em et al.} in \emph{A survey of named entity recognition and classification}~\cite{Nadeau07} point out three factors that distinguish the NERC algorithms: the language, the textual genre or domain, and the entity type. Regarding the third one, based on the Grishman {\em et al.} definition~\cite{Grishman96}, named entity refers to the name of a person or an organization, a location, a brand, a product, a numeric expression (including time, date, money and percentage), found in a sentence, but generally, the most studied types consider the \emph{enamex} designation --- proper names of `persons', `locations' and `organizations' --- the `miscellaneous' category for the proper names that fall outside the classic \emph{enamex}). In recent research , the possible types to extract are open and include subcategories~\cite{Nadeau07}. 

The language is an important factor to be taken in consideration in the NER task. Most of the services are devoted to English and few support NER on Portuguese texts. The first reference to work developed in Portuguese texts was published in 1997~\cite{Palmer97}; the authors perform the NER task and compute some measures in a Portuguese corpus and other five corpora. Until now, we have only identified the Rembrandt tool as a service developed and devoted to extract named entities in Portuguese texts. Other tools (\emph{AlchemyAPI, NERD} and \emph{Zemanta}) have been adapted to work and accept Portuguese texts but were not specifically developed for that purpose. As recently pointed out by Taba and Caseli~\cite{Taba14}, the Portuguese language still lacks high quality linguistic resources and tools.

NER is not only one task of the text mining process but also an initial step in the performance of other tasks, such as relation extraction, classification and/or topic modelling~\cite{Campos12}. This makes the quality of the NER process particularly important. In the light of the related works and taking in consideration that most of the approaches optimize $precision$ but not $recall$, we propose PAMPO to extract named entities in Portuguese texts. In this work we do not classify neither disambiguate the entity. Our major concern is to increase the $recall$ without decreasing the $precision$ of the named entity extractor. 

\section{The entity extraction algorithm}
\label{sec2}

In this work, we consider the \emph{enamex} definition of entities plus the miscellaneous named entities where we include events like, for instance, \emph{`Jogos Ol\'impicos'} (\emph{`Olympic Games'}). To identify those entities, an information extraction procedure was designed using regular expressions and other pattern matching strategies, along with part-of-speech tagging, {\em i.e.}, employing a Part-of-Speech Tagger (POST) tool. The extraction of the named entities from Portuguese unstructured texts is composed of two phases: candidate generation, where we generate a superset of candidate entities, and entity selection, where only relevant candidates are kept. The two phases are described in Algorithms~\ref{algo1} and \ref{algo2}, respectively.

{\bf PAMPO - Candidate Generation}
In this phase, we provide a customizable base of regular expressions that gathers common candidate entities. Typical expressions capture capitalized words, personal titles (president, deputy, etc.) and other common words (assembly). This patterns' base is extendable and the aim of the process in this phase is to identify all good candidates.

\begin{algorithm}
\caption{\hspace{.2cm} PAMPO - Candidate Generation}
\label{algo1}

\begin{algorithmic}
\State {\bf Input:} $Text$, $TPB$:Term Pattern Base
\State $CE \gets \emptyset$ \Comment{$CE$ is the set of candidate entities}
\For{each sentence $s$ in $Text$}
	\For{each term pattern $tp$ in TPB}
		\State{$CE \gets  CE \cup $ sub-sequences of $s$ that match $tp$}
        \EndFor
\EndFor
\State {\bf Output:} $CE$
\end{algorithmic}
\end{algorithm}

{\bf PAMPO - Entity Selection}
Here, all candidate entities of the previous phase are part-of-speech tagged. The POST process tags tokens with their corresponding word type (lexical category). Based on the tagging of the terms in candidate entities, we can identify some that can be discarded. This is done by applying a second level of regular expressions. In the entity selection phase, the regular expressions are defined on the lexical categories instead of terms themselves. For example, if the first word type is a `pron-det' (POS tag meaning determiner pronoun) the word is removed. Another example is the removal of candidate entities that do not have at least one tag `prop' or `n' (POS tag meaning a proper noun and a noun).

\begin{algorithm}

\caption{\hspace{.2cm} PAMPO - Entity selection}
\label{algo2}

\begin{algorithmic}
\State {\bf Input:} $CE$: candidate entities, $CPB$: category clipping patterns, $PPB$: category pruning patterns, $TPPB$: term pruning pattern base
\For{each candidate entity $c$ in $CE$}
	\State $post_c \gets$ POST of the candidate entity $c$
	\For{each clipping pattern $cp$ in $CPB$}
		\If{$cp$ matches prefix of $post_c$}
			\State remove matching prefix from $post_c$
			\State remove corresponding prefix from $c$
		\EndIf
        \EndFor
	\For{each pruning pattern $pp$ in $PPB$}
		\If{$pp$ matches $post_c$}
			\State $CE \gets CE \setminus \{c\}$
		\EndIf
        \EndFor
	\For{each pruning pattern $tp$ in $TPPB$}
		\If{$c$ = $tp$}
			\State $CE \gets CE \setminus \{c\}$
		\EndIf
        \EndFor        
\EndFor
\State {\bf Output:} modified $CE$
\end{algorithmic}

\end{algorithm}


\subsection{Implementation}

The program was developed in {\ttfamily R}~\cite{Rpro} and makes use of some specific text mining packages. We have implemented our method using the following {\ttfamily R} packages: {\bf tm}~\cite{tm14}, {\bf cwhmisc}~\cite{cwhmisc14}, {\bf memoise}~\cite{memoise14}, {\bf openNLP}~\cite{opennlp14}, {\bf Hmisc}~\cite{hmisc14}. The OpenNLP POS Tagger uses a probability model to predict the correct POS tag and, for Portuguese language, it was trained on CoNLL\_X bosque data.

\subsection{An application}

\begin{figure}[ht]
\begin{flushleft}
 \mbox{"}
\end{flushleft}
\centering
\vspace{-.4cm}
	\includegraphics[width=4.7in]{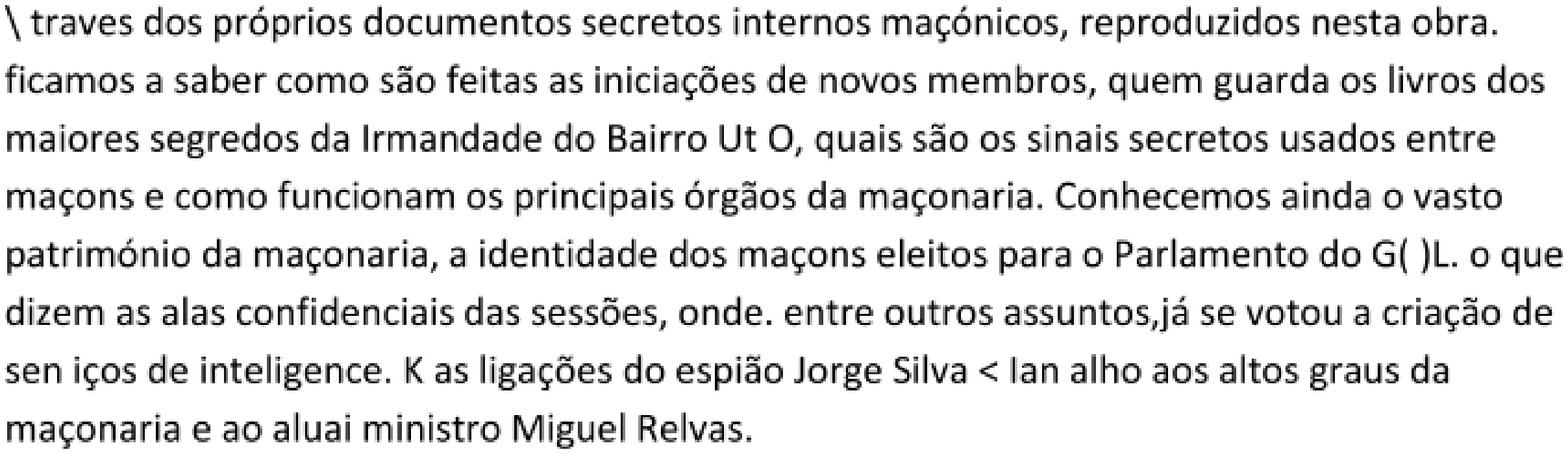}
	\begin{flushright}
	\vspace{-.7cm}
			 \mbox{"}
	\end{flushright}
\centering
\caption{A paragraph scanned from the book with OCR software.}
	\label{fig:livro}
\end{figure}

The $TPB$, $CPB$, $PPB$ and $TPPB$ bases adopted, for Portuguese texts, and used in this application are described in this section. As a first approach, and to test the PAMPO algorithm, we selected a book about the Portuguese Freemasonry~\cite{Vilela13}. Despite being on a specific topic, it contains a rich variety of situations to test our extractor.  
As an example, the piece of text shown in Figure~\ref{fig:livro} was scanned from the book with current OCR software and will be used here to highlight the contribution of each phase to the final result. The five named entities manually identified in this piece of text are \emph{`Irmandade do Bairro Ut O', `Parlamento do G', `Jorge Silva', `Ian'} and \emph{`ministro Miguel Relvas'}. 

Applying Algorithm~1 to the paragraph of Figure~\ref{fig:livro}, the set of `candidate entities' found are \emph{`Irmandade do Bairro Ut O', `Conhecemos', `Parlamento do G', `L', `K', `Jorge Silva', `Ian'} and \emph{`ministro Miguel Relvas'}. Although most of the words in the extracted `candidate entities' list start with capital letter, with this algorithm we were able to extract also other important words that are not capitalized like the first word in the last named entity (\emph{ministro}). This is possible because the $TPB$ base includes a set of patterns that captures not only words (or sequence of words) starting with capital letters but also words that are associated to some entity's name like the ones in list1 on Appendix A. 

Having collected the `candidate entities' in the previous step, we now proceed by removing from that list the ones that do not correspond to named entities. For that purpose, we use list2 (see Appendix A) as $CPB$ base, all the tags that are not a noun ($n$) or a proper noun ($prop$) are included in the $PPB$ base and, finally, some terms that are not named entities but that were not excluded by previous actions (see list3 on Appendix A), are used as $TPPB$ base. 
Applying Algorithm~2 with those lists to the set of `candidate entities', from Figure~\ref{fig:livro}, we obtain as named entities \emph{`Irmandade do Bairro Ut O', `Parlamento do G', `Jorge Silva', `Ian'} and \emph{`ministro Miguel Relvas'}. In fact, these five terms are the only named entities in the paragraph.

\subsection{Analysis of results}

Table~\ref{cloud} shows the most frequent `candidate entities' from the whole book, as extracted by Algorithm 1 and which of those candidate entities were considered as actual `named entities' by Algorithm 2. 

\begin{figure*}
\centering
			\includegraphics[width=2.3in]{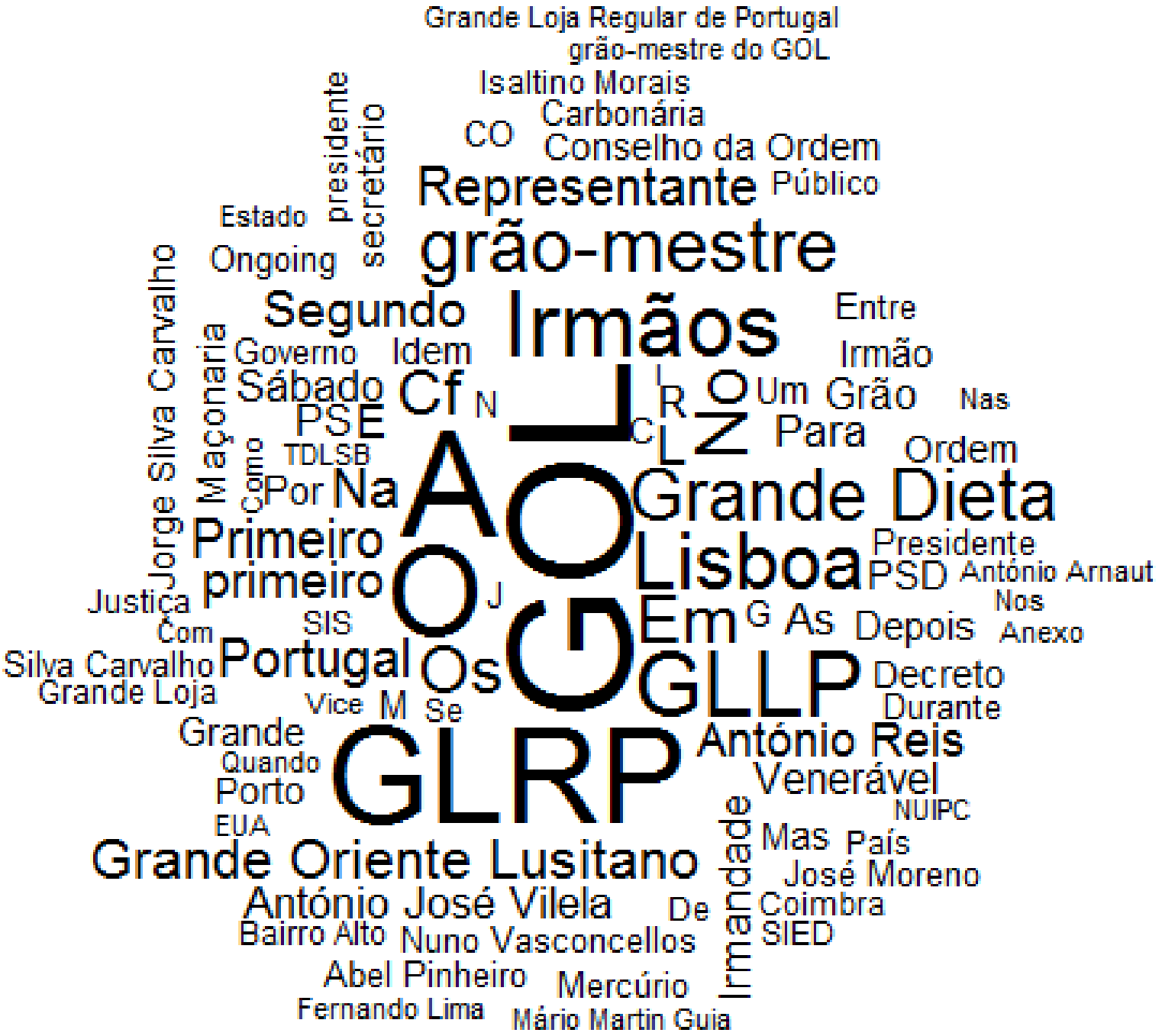}
			\includegraphics[width=2.3in]{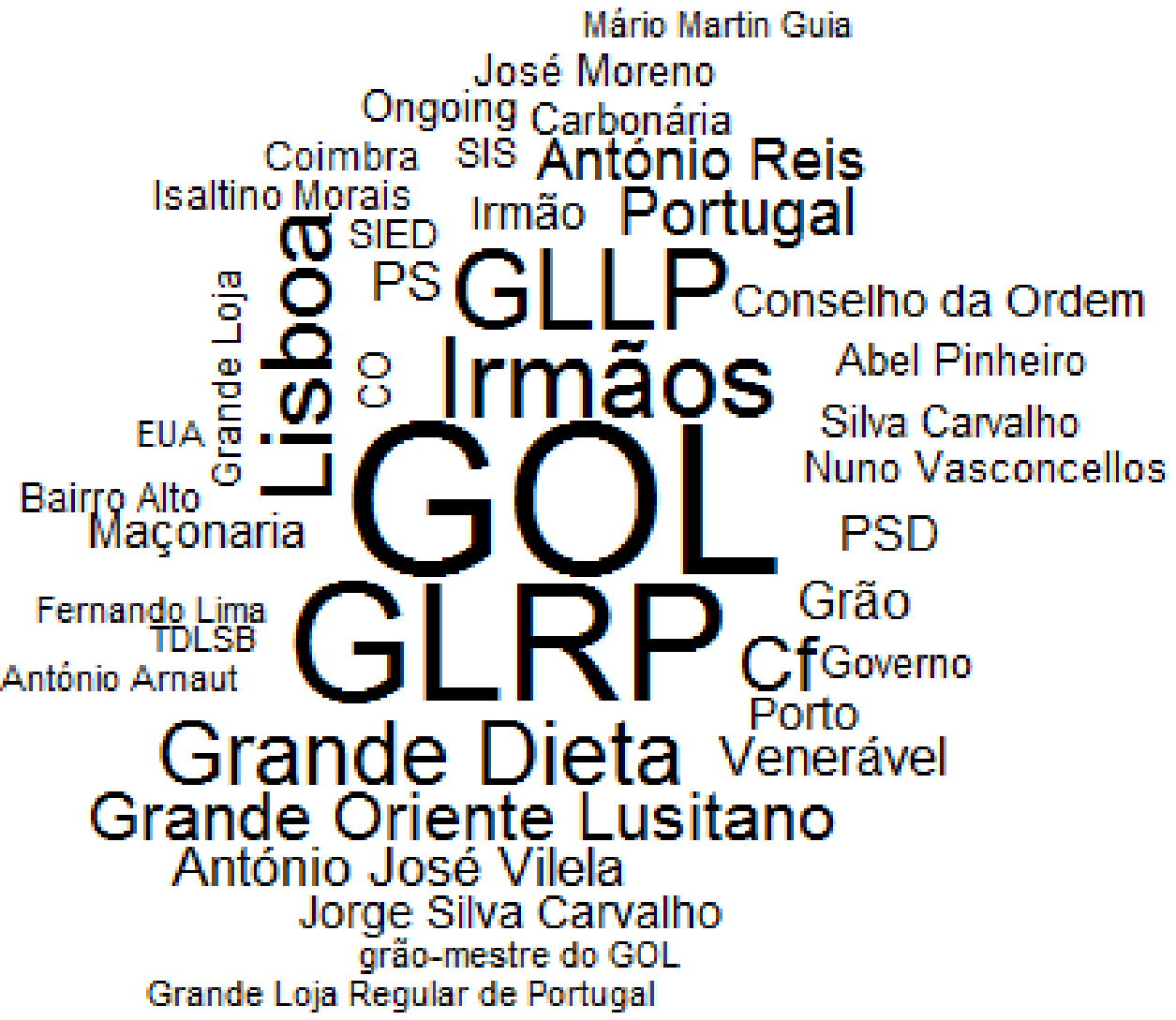}
			 \mbox{ \bf {\hspace*{ 0.1 in} Candidate Generation \hspace*{ 1.2in} Entity Selection\hspace*{ 0.3 in}}}
	\caption{Entity word clouds depicting the `candidate entities' that appear 25 or more times in the text. Each subfigure represents the set of `candidate entities' or`named entities' returned at the end of each phase of PAMPO process.}
	\label{fig:wordcloud}
\end{figure*}

To give an idea of the improvement introduced by each phase, we represent the `candidate entities' set in a word cloud where words with higher frequency have larger font size. As it can be observed in Figure~\ref{fig:wordcloud}, after phase~1 some words that do not refer to entities, such as \emph{`Idem'(`Idem'), `Entre' (`Between')} and \emph{`Nas' (`At the')}, are present in the cloud, but, as expected, they disappear in phase 2.

\begin{table}
\centering
\caption{ Some results of the application of PAMPO to the whole book. In the left column we have the most frequent `candidate entities' (please note that the column is folded in two), in the right column we have a `$+$' if the algorithm considers it a `named entity' and a `$-$' if not.}
\footnotesize
\begin{minipage}{5.5cm}
\vspace{.4cm}
\begin{tabular}{|l|c|} \hline\hline
{`candidate entity'}&{`named entity'} \\ \hline\hline
GOL& +\\
A&-\\
GLRP&+\\
O&-\\
Irm\~aos&+\\
GLLP&+\\
Lisboa&+\\
gr\~ao-mestre&-\\
Grande Dieta&+\\
No&-\\
Em&-\\
Os&-\\
Cf&+\\
L&-\\
Na&-\\
Grande Oriente Lusitano&+\\
Representante&-\\
Portugal&+\\
Primeiro&-\\
Segundo&-\\
\hline\hline
\end{tabular}
\end{minipage}
\begin{minipage}{5.5cm}
\begin{tabular}{|l|c|} 
\multicolumn{2}{l}{(Cont.)}\\\hline\hline
{`candidate entity'}&{`named entity'} \\ \hline\hline
primeiro&-\\
Ant\'onio Reis&+\\
E&-\\
Ant\'onio Jos\'e Vilela&+\\
R&-\\
S\'abado&-\\
As&-\\
Irmandade&-\\
Para&-\\
PS&+\\
Gr\~ao&+\\
Vener\'avel&+\\
M&-\\
Por&-\\
Ma\c{c}onaria&+\\
Grande&-\\
Depois&-\\
PSD&+\\
secret\'ario&-\\
Mas&-\\
\hline\hline
\end{tabular}
\end{minipage}
\label{cloud}
\end{table}

\begin{table}
\caption{Measures of the PAMPO process quality obtained for the 125 pages of the book.}
\footnotesize
\begin{tabular}{||l|c|c||} \hline\hline
Variable&Candidate Generation&Entity Selection\\ \hline\hline
extracted terms & \multirow{2}{*}{5089} & \multirow{2}{*}{3075}\\
(`candidate entity'/`named entity')&&\\ 
real named entities among the extracted terms & 3205& 2982\\
$recall$& 0.84& 0.78\\ 
$precision$ & 0.63& 0.97\\
$F_1$ & 0.72& 0.87\\
\hline\hline
\end{tabular}
\label{measure}
\end{table}

From this book, a total of 12120 named entities were extracted by PAMPO, corresponding to 5159 unique named entities. To assess the quality of this process, the first 125 pages of the book were manually labelled (1/3 of the text book). The values of the computed measures are shown in Table~\ref{measure}. This part of the book contains 3836 named entities. $Recall$ and $precision$ are estimated for the two phases based on  the results obtained on the 125 pages of the book. A total of 5089 terms were labelled `candidate entities' in the first phase and 3075 were identified as `named entities' in the second phase. The true positives were 3205 in the first phase and 2982 in the second phase (partial identifications count as 1/2). This means that the $recall$, given by Equation~(\ref{eq1}), decreases from 0.84 to 0.78,  and the $precision$,  given by Equation~(\ref{eq2}), increases from 0.63 to 0.97. 

\begin{equation}
\textstyle{
{  recall}=
\frac{\mbox{\footnotesize number of named entities extracted}}{\mbox{ \footnotesize number of named entities present in the text}}}
\label{eq1}
\end{equation}

\begin{equation}
\textstyle{
{precision}=
\frac{\mbox{ \footnotesize number of named entities extracted}}{\mbox{\footnotesize  number of `named entities' obtained with the process}}}
\label{eq2}
\end{equation}

Equation~(\ref{eq3}) defines another measure commonly used to assess the quality of the process, $F_1$. This measure allows interpreting the global quality, taking into account the decrease of $recall$ and the increase of $precision$. The second phase of the PAMPO process increases the value of $F_1$ from 0.72 to 0.87. 

\begin{equation}
\textstyle{
{F_1}=2 \times
\frac{\mbox{$ precision\times recall$}}{\mbox{$ precision+recall$}}}
\label{eq3}
\end{equation}

After these illustrative results of the PAMPO algorithm, the following section presents the results of a comparison between PAMPO and other approaches to extract named entities from texts in Portuguese.

\section{Comparing PAMPO with other NER tools}

\begin{table}
  \caption{Corpus Information}
    \begin{tabular}{c|c|rr}\hline\hline
   \multicolumn{2}{c|}{}& News & Sports news \\ \hline \hline
  \multicolumn{2}{c|}{Documents}& 227&881\\\hline
	\multirow{3}{3cm}{Number of Words by document} &minimum & 24&59\\\cline{2-4}
	&maximum & 770&445\\\cline{2-4}
	&mean & 263.4&169.3\\\hline
  \multicolumn{2}{c|}{Entities}& 3671&14516\\\hline
	\multirow{4}{4cm}{Type of entities} &Person & 1195&7051\\\cline{2-4}
	&Location & 831&3285\\\cline{2-4}
	&Organization & 847&1017\\\cline{2-4}
	&Miscellaneous& 798&3163\\\hline\hline
    \end{tabular}%
  \label{tab:addlabel0}%
\end{table}%

In this work, we evaluate our NER approach using two news corpora. One corpus is a set of 227 texts published on December 31, 2010 by the \emph{Lusa agency} (portuguese agency of news) and will be referred to as `News'. The other corpus (named here `Sports news') is a set of 881 sports news\footnote{Described and available at {`http://www.researchgate.net/publication/264789916\_bestsports-v1.0'}}. The texts were manually annotated according to the \emph{enamex} designation and the type `miscellaneous'. 

Each of the corpora used for evaluation has a considerable number of texts but with different characteristics. The `Sports news' corpus has text from only one domain, while the `News' presents a diversity of topics. This fact allows evaluating if the domain/topic factor can significantly affect the quality of the algorithm. Some features of the two corpora are present in Table~\ref{tab:addlabel0}. The minimum text length in words is 24 for the `News' corpus and 59 for `Sports news'. The maximum lengths are 770 and 445 respectively. The total named entities manually found for each type range between 798 and 7051 with an average of 16.4 entities (without type distinction) per text.

In this work we not only study the quality of the PAMPO NER extractor for Portuguese texts but we also compare the results with three other extractors. Two of them, AlchemyAPI and Zemanta, are easily accessed with the tool developed by Bartosz Malocha in EURECOM and available on the web\footnote{http:// entityextraction.appspot.com/}. The other one, Rembrandt, has to be downloaded and locally installed, which is not a straightforward task.

\subsection{PAMPO output}

Considering the Portuguese text represented in Figure~\ref{texto_sport} (a) the PAMPO algorithm identifies the `named entities' listed in Figure~\ref{texto_sport} (b). 

\begin{figure*}
\begin{minipage}[t]{7.5cm}
\footnotesize
\emph{{\bf Brasil} levar\'a delega\c{c}\~{a}o recorde aos {\bf Jogos Ol\'impicos de Atenas 2004}}

\vspace{.5cm}
\emph{A delega\c{c}\~{a}o brasileira nos {\bf Jogos Ol\'impicos de Atenas 2004} ser\'a a maior da hist\'oria. 
Ap\'os os resultados do {\bf Trof\'eu Brasil de Atletismo}, no \'ultimo final de semana, em {\bf S\~{a}o Paulo}, a delega\c{c}\~{a}o brasileira atingiu um total de 234 atletas, superando a marca anterior de 225 atletas nos {\bf Jogos Ol\'impicos de Atlanta 1996}. }

\emph{At\'e o momento s\~{a}o 118 atletas homens e 116 atletas mulheres. E este n\'umero ainda pode crescer. O {\bf COB} aguarda a defini\c{c}\~{a}o do t\^{e}nis, cujo an\'uncio, de acordo com o ranking, acontecer\'a na pr\'oxima semana pela {\bf Federa\c{c}\~{a}o Internacional de T\^{e}nis}. Isso ocorrendo, o {\bf Brasil} ter\'a a participa\c{c}\~{a}o em 26 esportes. O recorde anterior era de 24 esportes, em {\bf Sydney 2000}. }

\emph{Na nata\c{c}\~{a}o tamb\'em h\'a a expectativa de o {\bf Brasil} classificar os revezamentos femininos pelo ranking da {\bf Federa\c{c}\~{a}o Internacional de Nata\c{c}\~{a}o}. H\'a tamb\'em chances no remo, cuja \'ultima seletiva acontece a partir de domingo, em {\bf Lucerne}, {\bf Su\'i\c{c}a}. }

\emph{`Esse recorde demonstra a evolu\c{c}\~{a}o qualitativa do esporte ol\'impico brasileiro. Isso vem ocorrendo desde os {\bf Jogos Ol\'impicos de Atlanta 1996}, gra\c{c}as ao trabalho que vem sendo feito pelas {\bf Confedera\c{c}\~{o}es Brasileiras Ol\'impicas} em conjunto com o {\bf COB}. Com a {\bf Lei Agnelo/Piva}, essa evolu\c{c}\~{a}o vem sendo ainda mais efetiva, j\'a que estamos podendo fazer um planejamento e execut\'a-lo de forma cont\'inua. Vale ressaltar que todos os pa\'ises tamb\'em est\~{a}o evoluindo. Estou muito feliz com esse recorde, at\'e porque nenhum atleta est\'a indo por meio de convite. Todos garantiram as vagas pelos crit\'erios t\'ecnicos de suas {\bf Federa\c{c}\~{o}es Internacionais}', afirmou o {\bf presidente do COB}, {\bf Carlos Arthur Nuzman}.}
\end{minipage} 
\hfill
\vline
\hfill
\begin{minipage}[t]{4.2cm}
{\bf `named entity'}\\
\vspace{.2cm}

\footnotesize
Brasil\\
Jogos Ol\'impicos de Atenas\\
Jogos Ol\'impicos de Atenas\\
Trof\'eu Brasil de Atletismo\\
S\~{a}o Paulo\\
Jogos Ol\'impicos de Atlanta\\
COB\\
Federa\c{c}\~ao Internacional de T\^enis\\
Brasil\\
Sydney\\
Brasil\\
Federa\c{c}\~ao Internacional de Nata\c{c}\~ao\\
Lucerne\\
Su\'i\c{c}a\\
Jogos Ol\'impicos de Atlanta\\
Confedera\c{c}\~oes Brasileiras Ol\'impicas\\
COB\\
Lei Agnelo\\
Piva\\
Vale\\
Federa\c{c}\~oes Internacionais\\
presidente do COB\\
Carlos Arthur Nuzman
\end{minipage} 
	 \raisebox{-3ex}{ \bf {\hspace*{ 1.5 in} (a) \hspace*{ 1.5in} (b) \hspace*{ 0.3 in}}}
 \caption{An example of a text present in the `Sports news' corpus (a) and the list of `named entities' obtained by the PAMPO extractor in (b).}
\label{texto_sport}
\end{figure*}

 \raisebox{0pt}[0pt][0pt]{}
As can be observed by this example, the algorithm extracts all the manifestations of `named entities' and lists them in the order they appear in the text, including repetitions of the same `named entity'.

\subsection{Evaluation}
To compare the results of PAMPO with the other NER extractors, we compute the $precision$ and $recall$ considering a unique occurrence per entity, instead of all named entities occurrences. Figure~\ref{output4} presents the outputs of the four extractors, PAMPO, \emph{AlchemyAPI}, \emph{Rembrandt} and \emph{Zemanta}, for the text in Figure~\ref{texto_sport} (a).

\begin{figure*}
\centering
{\bf `named entities' lists}\\

\vspace{.5cm}
\footnotesize
\begin{minipage}[t]{5.5cm}
{\bf PAMPO}\\

Brasil\\
Jogos Ol\'impicos de Atenas\\
Trof\'eu Brasil de Atletismo\\
S\~{a}o Paulo\\
Jogos Ol\'impicos de Atlanta\\
COB\\
Federa\c{c}\~{a}o Internacional de T\^{e}nis\\
Sydney\\
Federa\c{c}\~{a}o Internacional de Nata\c{c}\~{a}o\\
Lucerne\\
Su\'i\c{c}a\\
Confedera\c{c}\~{o}es Brasileiras Ol\'impicas\\
Lei Agnelo\\
Piva\\
Vale\\
Federa\c{c}\~{o}es Internacionais\\
presidente do COB\\
Carlos Arthur Nuzman
\end{minipage}
\hfill
\vline
\hfill
\begin{minipage}[t]{5cm}
{\bf Alchemy}\\

Brasil\\
Jogos Ol\'impicos\\
COB\\
Atlanta\\
t\^{e}nis\\
Atenas\\
Atletismo\\
ol\'impico\\
S\~{a}o Paulo\\
Federa\c{c}\~{o}es Internacionais\\
Sydney\\
Carlos Arthur Nuzman\\
Federa\c{c}\~{a}o Internacional de T\^{e}nis
\end{minipage}
\vspace{.3cm}

\hrule

\vspace{.3cm}

\begin{minipage}[t]{5.5cm}
{\bf Rembrandt}\\

Brasil \\
Jogos \\
Atenas \\
Paulo \\
Atlanta \\
Sydney \\ 
Lei\_Agnelo\_/\_Piva \\
presidente\_do\_COB \\
Carlos\_Arthur\_Nuzman 
\end{minipage} 
\hfill
\vline
\hfill
\begin{minipage}[t]{5cm}
{\bf Zemanta}\\

Carlos Arthur Nuzman\\
Brasil\\
Atlanta 1996\\
Internacional\\
Sydney 2000\\
S\~{a}o Paulo\\
Lucerne
\end{minipage} 
 \caption{Lists of `named entities' obtained by each of the four extractors, PAMPO, \emph{AlchemyAPI}, \emph{Rembrandt} and \emph{Zemanta}, for the Portuguese text represented in Figure~\ref{texto_sport} (a).}
\label{output4}
\end{figure*}

\begin{table}
  \caption{Summary statistics of extractors'performance}
    \begin{tabular}{crrrrr}\hline\hline
         &  & PAMPO & Alchemy & Rembrandt& Zemanta \\ \hline\hline 
   \multirow{4}{1.5cm}{Recall} & \multirow{2}{1.5cm}{News}& 0.910 & 0.465 & 0.456& 0.356 \\ 
		&& \tiny(0.094)  & \tiny(0.167) &\tiny (0.159) & \tiny(0.166)\\ \cline{2-6}
	&\multirow{2}{1.5cm}{Sports news}	&  0.959 & 0.675 & 0.587 & 0.499\\ 
		&& \tiny(0.054) & \tiny(0.127) &\tiny (0.175) & \tiny(0.187) \\
		\hline\hline
    \multirow{4}{1.5cm}{ Precision}&\multirow{2}{1.5cm}{News} & 0.964  & 0.947 & 0.846 & 0.841\\
		&&\tiny(0.063)& \tiny(0.147) & \tiny(0.223) & \tiny(0.210)\\ \cline{2-6}
		 &\multirow{2}{1.5cm}{Sports news}	& 0.984  & 0.984 & 0.936 & 0.921\\ 
		&& \tiny(0.036)  & \tiny(0.055) & \tiny(0.109)& \tiny(0.136)\\ \hline\hline
   \multirow{4}{1.5cm}{$F_1$}&\multirow{2}{1.5cm}{News} & 0.932  & 0.608 & 0.561 &  0.481 \\  
		&& \tiny(0.066)  & \tiny(0.168) & \tiny(0.162) & \tiny(0.180)\\\cline{2-6}
		&\multirow{2}{1.5cm}{Sports news}	& 0.971 & 0.794 & 0.706& 0.628  \\
		&& \tiny(0.036)  & \tiny(0.098) & \tiny(0.152)  & \tiny(0.181)\\ \hline\hline
   
    \end{tabular}%
  \label{tab:rpf}%
\end{table}%

To compute the $recall$, $precision$ and $F_1$ measures presented in Table~\ref{tab:rpf}, we used Equations~\ref{eq1}, \ref{eq2} and \ref{eq3} with a difference in the weight given to the partial identifications. Based on the example in Figure~\ref{output4}, we observed that not all partial correspondences to the named entity on the text have necessarily the same value, {\em i.e.}, \emph{`Atlanta'}, \emph{`Atlanta 1996'}, \emph{`Jogos Ol\'impicos'} or \emph{`Jogos Ol\'impicos de Atlanta'} as partial identifications of \emph{`Jogos Ol\'impicos de Atlanta 1996'} do not have the same information. Hence we adopted as weight criterion for the partial identifications, the fraction of the named entity that is identified. This means that the previous partial identifications have weights of $1/4$, $2/4$, $2/4$ and $3/4$, respectively. As a result, two extractors will have the same performance even if one identifies the complete named entity \emph{`Jogos Ol\'impicos de Atlanta 1996'} and the other splits it into two named entities, \emph{`Atlanta 1996'} and \emph{`Jogos Ol\'impicos'}.

\begin{figure*}
\centering
			\includegraphics[width=4.in]{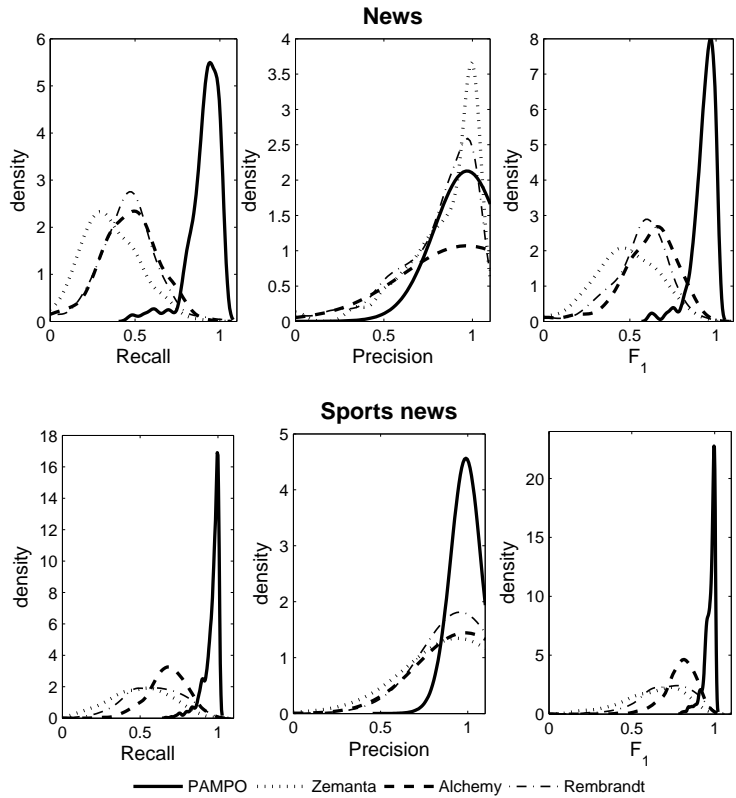}

	\caption{Estimated probability density function of the three measures, $recall$, $precision$ and $F_1$, for the two corpora.}
	\label{fig:density}
\end{figure*}

Analysing the mean values of $recall$, $precision$ and $F_1$ (standard deviation between parentheses) given in Table~\ref{tab:rpf}, it is easy to conclude that they are higher in the `Sports news'  for all the extractors. Moreover, that difference is less noted in the PAMPO algorithm, which presents better results and a much higher mean $recall$, and consequently higher mean $F_1$, than the other three extractors. The four extractors have similar mean $precision$ but none has better mean $precision$ than the PAMPO extractor. The mean $recall$, mean $precision$ and mean $F_1$ for the PAMPO algorithm are consistent with a good performance of the extractor. To further assess the quality of the extractors, the probability density function of the three measures for the two corpora, estimated using a kernel density estimation with 100 equally spaced points (MATLAB 7.10.0 (R2010a)), are plotted in Figure~\ref{fig:density}. As expected, the probability density is higher around the value 1 for all the measures of PAMPO extractor on the two corpora. 

\begin{figure*}
\centering
			\includegraphics[width=4.in]{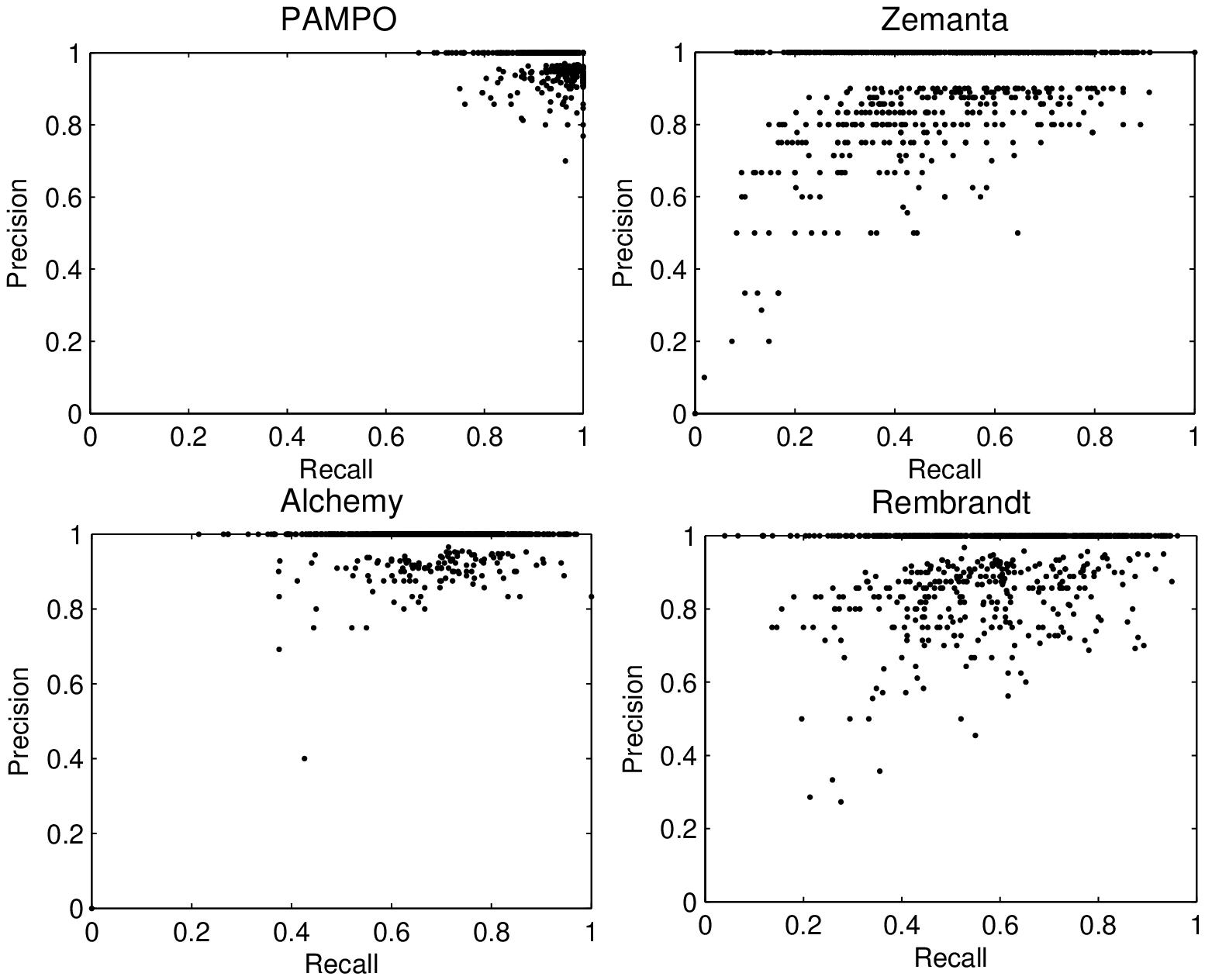}
			\includegraphics[width=4.in]{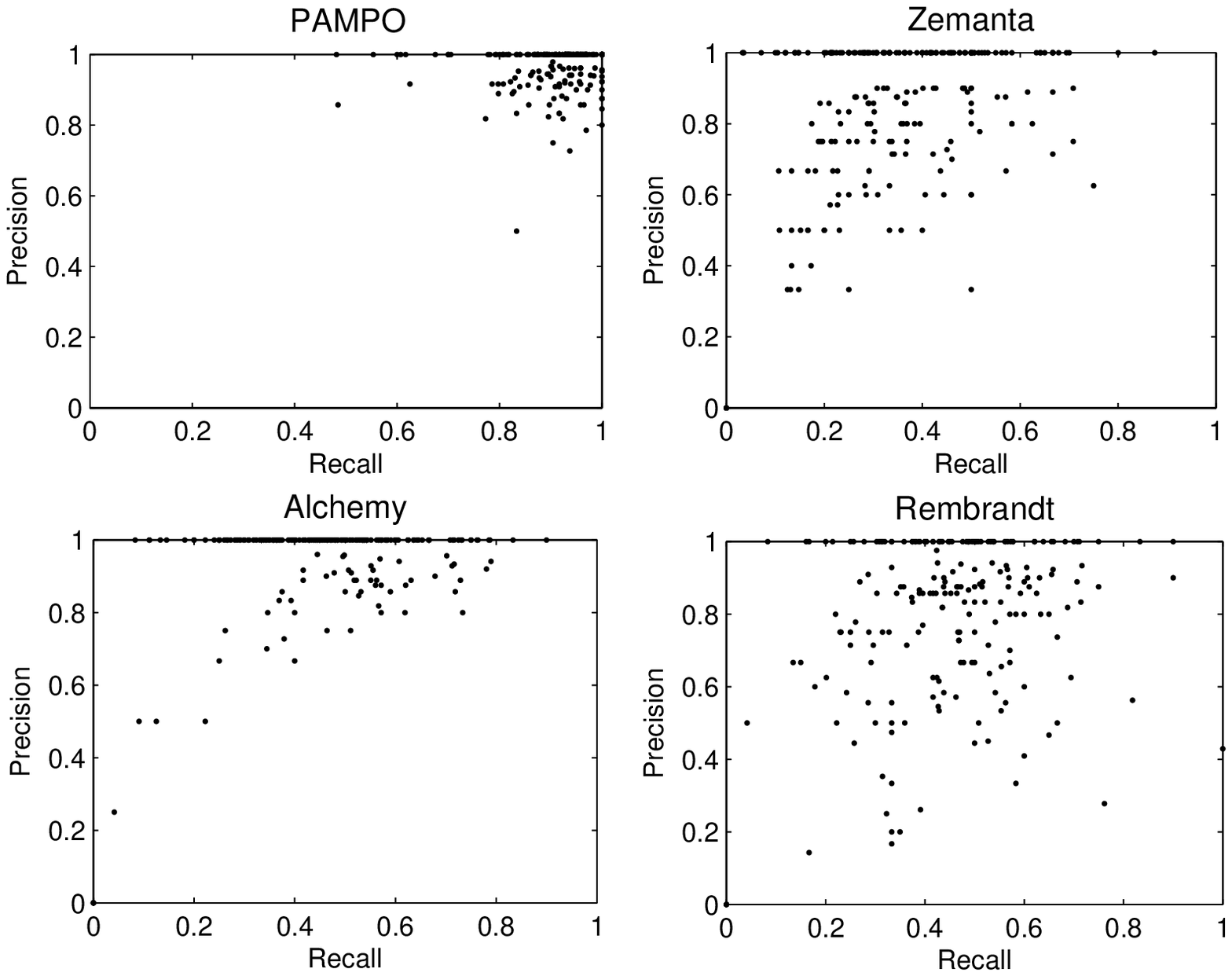}
	\caption{Scatter plots of $precision$ and $recall$ for four extractors, PAMPO, \emph{Alchemy}, \emph{Rembrandt} and \emph{Zemanta}, for the `Sports news' in the first four panels and for the news published by \emph{Lusa} agency in the four bottom panels.}
	\label{fig:pvsr_desp}
\end{figure*}

Figure~\ref{fig:pvsr_desp} presents scatter plots of $precision$ {\em vs} $recall$ for the four extractors, PAMPO, \emph{AlchemyAPI}, \emph{Rembrandt} and \emph{Zemanta} for the `Sports news' and `News' corpora, first four panels and four bottom panels, respectively. It is noteworthy that almost all the 881 points of the `Sports news'  for PAMPO extractor are in the upper right corner of the scatter plot, as well as almost all the 227 points of the `News'. The other tools present  a more dispersed solution quality.

\subsection{Evaluation by type of entity}

To determine if the entity type contributes to output variability in the $recall$, an analysis was conducted on the named entities for the classification types: `persons' (PER), `locations' (LOC), `organizations' (ORG) and `miscellaneous' (MISC).

\begin{figure*}
\centering
			\includegraphics[width=4.5in]{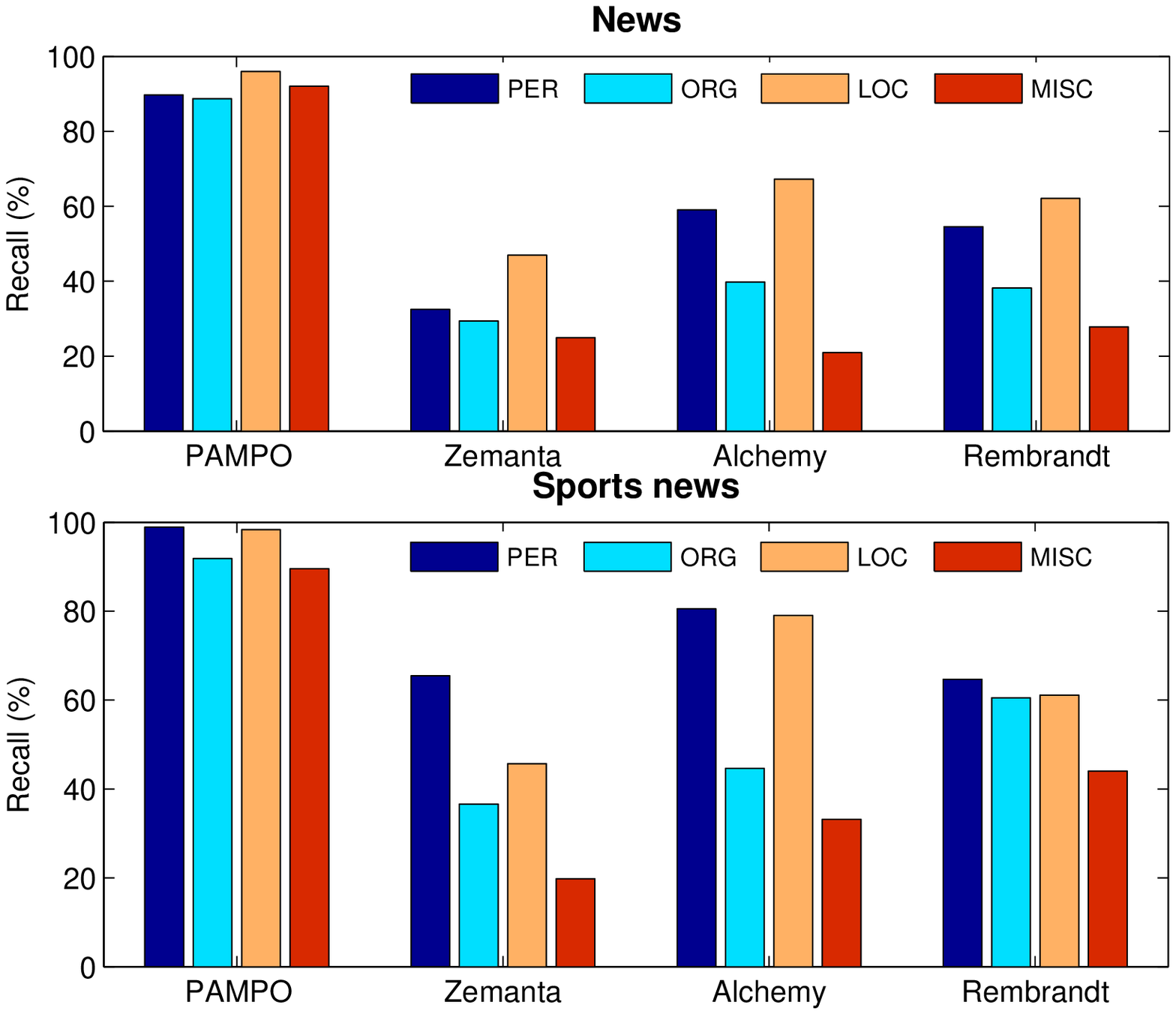}

	\caption{Bar representation for the $recall$ of the four extractors, PAMPO, \emph{Alchemy}, \emph{Rembrandt} and \emph{Zemanta}, for the four types of entities, `persons' (PER), `locations' (LOC), `organizations' (ORG) and `miscellaneous' (MISC), and for the two corpora.}
	\label{fig:recall}
\end{figure*}

The results (Figure~\ref{fig:recall}) indicate that the $recall$ varies with the type of entity for the  \emph{AlchemyAPI}, \emph{Rembrandt} and \emph{Zemanta} but not for the PAMPO. The $recall$ of PAMPO extractor is the highest for all types of entities.

In summary, it is apparent from the analysis that PAMPO extracts a set of `named entities' that resembles the actual list of named entities on texts.

To complete the evaluation we also computed $precision$, $recall$ and $F_1$ of PAMPO extraction on the texts in Cole\c c\~ao Dourada-\emph{HAREM}~\footnote{http://www.linguateca.pt/HAREM/}. This corpus has 129 documents. Using the evaluation criterion defined by curators of \emph{HAREM}, we obtain a $precision$ of $0.788$, a $recall$ of $0.722$ and a $F_1$ of $0.736$ considering all the categories. Considering that the PAMPO extractor was not designed to extract quantities or time expressions we computed the same measures excluding these two types of entities. While $precision$ practically keeps the same value ($0.787$), $recall$ and $F_1$ increase to $0.894$ and $0.837$, respectively.

\subsection{PAMPO versus three other extractors}

Now, we analyse the differences between measures obtained with PAMPO and  with the three other extractors, for each one of the news on the two corpora. To perform a more informative comparison between PAMPO and the other extractors, we count the number of news items that had a positive, a null and a negative difference with respect to each measure and each concurrent extractor. These are summarized in Table~\ref{tab:addlabel1} for both corpora.

\begin{table}
  \centering
  \caption{Number of positive and negative occurrences in the difference between the $recall$, $precision$, and $F_1$ of PAMPO and the three other extractors, \emph{AlchemyAPI}, \emph{Rembrandt} and \emph{Zemanta}, for the two corpora, `Sports news' and `News'} \label{tab:addlabel1}
    \begin{tabular}{cc|rrr||rrr}
			\hline\hline
		\multicolumn{2}{c|}{} &\multicolumn{3}{c||}{Sports news} &\multicolumn{3}{c}{News} \\
    \multicolumn{2}{c|}{} & Alchemy & Rembrandt & Zemanta  &  Alchemy & Rembrandt & Zemanta\\ \hline\hline
     \multirow{3}{1.cm}{Recall} & diff $>0$  & 227   & 225& 227  &  877   & 878&878  \\
		& diff $=0$  & 0     & 2&  0    & 3     & 2  &1    \\
    & diff $<0$& 0     & 0     & 0 &     1     & 1   & 2    \\ \hline
 \multirow{3}{1.cm}{Precision} &diff $>0$  & 51    & 146 &117   & 123   & 300& 318   \\
& diff $=0$   & 110   & 50   &74   & 603   & 474  & 451  \\
&diff $<0$   & 66    & 31  & 36  & 155   & 107  &112\\\hline     
\multirow{3}{1.cm}{$F_1$} &diff $>0$  & 225   & 226 &227    &875   & 880 &879  \\
& diff $=0$    & 1     & 0   &0 &2     & 0    & 0  \\
& diff $<0$    & 1     & 1    & 0  & 4     & 1  &2    \\ \hline\hline
    \end{tabular}
\end{table}

The mean and the standard deviation (between parentheses) for each extractor and each corpus are presented in Table~\ref{tab:addlabel3}. They will be used to test statistical hypotheses about the mean difference value of $recall$, $precision$ and $F_1$ between PAMPO and the other three extractors.

\begin{table}
  \centering
  \caption{Summary statistics of the difference between the performance of the PAMPO extractor and the other three extractors, \emph{AlchemyAPI}, \emph{Rembrandt} and \emph{Zemanta}, for the two corpora, `Sports news' and `News'.}
    \begin{tabular}{ccrrr}\hline\hline
    \multicolumn{2}{c}{} & Alchemy & Rembrandt& Zemanta   \\ \hline\hline
     \multirow{4}{1.2cm}{Recall} &  \multirow{2}{1.5cm}{News} & 0.445 &0.453 &  0.554 \\
		&  & \tiny (0.172) & \tiny (0.168) & \tiny (0.166) \\\cline{2-5}
		& \multirow{2}{1.5cm}{Sports news}  & 0.284 & 0.372 & 0.460\\
		&  & \tiny (0.127) &\tiny ( 0.175) &\tiny (0.185)  \\ \hline
 \multirow{4}{1.2cm}{Precision} & \multirow{2}{1.5cm}{News} & 0.017 & 0.159 &  0.123\\
&& \tiny (0.156) & \tiny (0.233)  & \tiny (0.210)  \\\cline{2-5}
& \multirow{2}{1.5cm}{ Sports news} & 0.001 & 0.048&0.063 \\
& & \tiny (0.066) & \tiny (0.111) &\tiny (0.138)  \\\hline
\multirow{4}{1.2cm}{$F_1$} &\multirow{2}{1.5cm}{News} &0.325 & 0.372 &0.452 \\
& & \tiny (0.170) &\tiny (0.172) &\tiny (0.177) \\\cline{2-5}
 & \multirow{2}{1.5cm}{Sports news}&0.177 & 0.265& 0.342 \\
& & \tiny (0.098) & \tiny (0.150) &\tiny (0.175)\\ \hline\hline
    \end{tabular}%
  \label{tab:addlabel3}%
\end{table}%

Based on all the values of the differences between PAMPO and the other extractors, represented in Tables~\ref{tab:addlabel1} and~\ref{tab:addlabel3}, we may say that:

\begin{itemize}
\item the $recall$ of the PAMPO extractor is the highest in almost all the news;
\item $precision$ does not differ much between PAMPO and the other extractors;
\item as a consequence the $F_1$ of PAMPO is also the highest in almost all the news;
\item the mean difference of $recall$ between PAMPO and \emph{AlchemyAPI} seams to be at least greater than 0.25;
\item the mean difference of $recall$ between PAMPO and \emph{Rembrandt} seams to be at least greater than 0.35;
\item the mean difference of $recall$ between PAMPO and \emph{Zemanta} seams to be at least greater than 0.40;
\item the mean difference of $precision$ is positive but near zero for all the three extractors;
\item the mean difference of $F_1$ between PAMPO and \emph{AlchemyAPI} seams to be at least greater than 0.15;
\item the mean difference of $F_1$ between PAMPO and \emph{Rembrandt} seams to be at least greater than 0.25;
\item the mean difference of $F_1$ between PAMPO and \emph{Zemanta} seams to be at least greater than 0.30.
\end{itemize}

To test the null hypothesis that the mean $recall$ differences between PAMPO and the other extractors are equal to 0.25, 0.35 and 0.40, for \emph{AlchemyAPI}, \emph{Rembrandt} and \emph{Zemanta}, respectively, \emph{ztest} was performed considering as alternative the mean $recall$ differences greater than those values. Based on the results of these two corpora the p-values are smaller than 9.5E-05. Hence, the results obtained so far provide statistical evidence that PAMPO increases NER $recall$ by at least 0.25.

\section{Remarks and Conclusions}
\label{sec4}

In this work we propose a novel effective method to extract named entities from unstructured text. The proposed PAMPO method is implemented using free software, namely {\ttfamily R} and available packages. Two manually annotated Portuguese news corpora  were used to empirically evaluate the algorithm using the measures of $recall$, $precision$ and $F_1$. These corpora did not influence the definition of the algorithm or the construction of its pattern bases. We have compared PAMPO with three other NER extractors: \emph{AlchemyAPI}, \emph{Rembrandt} and \emph{Zemanta}. Experimental results clearly show that PAMPO obtains significantly higher $recall$ and $F_1$ than existing tools. The values of $precision$  are identical. We may say also that PAMPO's performance in the \emph{HAREM} corpus was at least as good as the best one of the systems reported over there when we consider all categories of entities. However, when we exclude dates and numeric expressions, it  presents better results than the ones reported for other tools.

Despite its simplicity, PAMPO has a very good performance and is highly configurable. The PAMPO algorithm is potentially adaptable to be used for other languages by properly defining the pattern bases. Furthermore, it allows for straightforward improvement of the results by adding terms to the lists.

The results take us one step closer to the creation of a text intelligence system to be used in several applications, namely, in the study of the social context of possible economic and financial offenses.
As future work the authors are planning to improve the text mining procedure, by including a classification and a disambiguation step, as well as by automatically characterizing the relations between entities.

\section*{Acknowledgements}
The authors would like to thank SAPO Labs (http://labs.sapo.pt) for providing the data set of news from \emph{Lusa} agency. 
The authors would also like to thank grant \#2014/08996-0 and grant \#2013/14757-6, S\~ao Paulo Research Foundation (FAPESP).
This work is partially funded by FCT/MEC through PIDDAC and ERDF/ON2
within project NORTE-07-0124-FE\-DER-000059 and through the COMPETE
Programme (operational pro\-gramme for com\-pe\-ti\-ti\-ve\-ness) and by National
Funds through the FCT - Funda\c{c}\~ao para a Ci\^encia e a Tecnologia
(Portuguese Foundation for Science and Technology) within project
FCOMP-01-0124-FEDER-037281.


\section*{Appendix}
\label{listas}

{\bf list1} - \{'gr\~{a}o$-$mestre', 'papa', 'duque', 'duquesa', 'conde', 'condessa', 'visconde', 'viscondessa', 'rei', 'ra\'inha', 'pr\'incipe', 'princesa', 'marqu\^es', 'marquesa', 'bar\~ao', 'baronesa', 'bispo', 'presidente', 'secret\'ario', 'secret\'aria', 'ministro', 'ministra', 'pri\-mei\-ro', 'primeira', 'deputado', 'deputada', 'general', 'tenente', 'capit\~ao', 'capit\~a', 'sargento', 'governador', 'governadora', 'diretor', 'director', 'diretora', 'directora', 'ex', 'filho', 'filha', irm\~ao', 'irm\~a', 'pai', 'm\~ae', 'tio', 'tia', 'padrinho', 'madrinha', 'sobrinho', 'sobrinha', 'afilhado', 'afilhada', 'av\'o', 'av\^o', 'neto', 'neta', 'enteado', 'enteada', 'padrasto', 'madrasta'\}\\

{\bf list2} - \{'pron-det', 'adv adv ', 'adv prop', 'adv adj ', 'adv v-fi'\}\\

{\bf list3} - \{'Aproveitamento', 'Cuidado', 'Decerto', 'Desta', 'Desenvolvimento', 'Lan\-\c{c}a\-men\-to', 'Levantamento', 'Muitos', 'Muitas', 'Nessa', 'Nesse', 'Nessas', 'Nesses', 'Nestes', 'Neste', 'Nesta', 'Nestas', 'Noutro', 'Outros', 'Outro', 'Outra', 'Outras', 'Onde', 'Poucos', 'Poucas', 'Perante', 'Pela', 'Rec\'em', 'Tal', 'V\'arios', 'V\'arias', 'V\'os', 'Aceite', 'Comprometo', 'Cabe', 'Coloca', 'Conhecemos', 'Casado', 'Considerava', 'Desejo', 'Dev\'iamos', 'Escolhiam, 'Executa', 'Faça', 'Fica', 'Interrompidas', 'Indicar', 'Inclu\'ido', 'Leva', 'Morrer', 'Ouvistes', 'Prestaste', 'Praticou', 'Pressiona', 'Pensa', 'Poder', 'Podes', 'Revolta', 'Sabe', 'Ser', 'Ter', 'Toque', 'Toma', 'Trata', 'Vens', 'Ve\-ri\-fi\-cou', 'Viver', 'Vivemos', 'Venho', 'Rea\c{c}\~{a}o', 'Sess\~ao', 'Testamento', 'Toler\^ancia', 'T\'ermino', 'Vit\'oria', 'Visita', 'Harmonia', 'Iniciado', 'Instala\c{c}\~ao', 'Ibidem', 'In\-ven\-ta\-ri\-a\-\c{c}\~ao', 'Irregularidades', 'Internet', 'Lda', 'Manuten\c{c}\~ao', 'Nomeado', 'Obedi\^encia', 'Peti\c{c}\~ao', 'Passaporte', 'Proposta', 'Programa', 'Proibi\c{c}\~ao', 'Paz', 'Publica\c{c}\~ao', 'Question\'ario', 'Quadro', 'Relat\'orio', 'Redu\c{c}\~ao', 'Reorganiza\c{c}\~ao','Revolu\c{c}\~ao', 'Rep\'ublica', 'Reequil\'ibrio', 'Anexo', 'Abertura', 'Atestado', 'Ata', 'Ado\c{c}\~ao', 'Atualiza\c{c}\~ao', '\`As', '\'A', 'Capa', 'Convite', 'Compromisso', 'Condecora\c{c}\~ao', 'Convocat\'oria', 'Cart\~ao', 'Causa', 'Comunica\c{c}\~ao', 'Corrup\c{c}\~ao', 'Converg\^encia', 'Decreto', 'Ditadura', 'De\-mo\-cra\-ci\-a', 'Democrata', 'Estrutura', 'Ficha', 'Fax', 'Fixa\c{c}\~ao', 'Futuro', 'Gabinete', 'Gl\'oria', 'Janeiro', 'Fe\-ve\-rei\-ro', 'Mar\c{c}o', 'Abril', 'Maio', 'Junho', 'Julho', 'Agosto', 'Setembro', 'Outubro', 'Novembro', 'Dezembro', Di\'ario', 'Semanal', 'Mensal', 'Minutos', 'Meses', 'Ano', 'Anos', 'Hoje'\}$\cup$\{Portuguese stopwords on {\ttfamily R}\}



\end{document}